\documentclass[aip,pop,reprint,citeautoscript]{revtex4-2}

\usepackage[colorlinks,urlcolor=blue,citecolor=blue]{hyperref}
\usepackage{color}
\usepackage[export]{adjustbox}
\usepackage{graphicx}
\usepackage{siunitx}
\DeclareSIUnit\angstrom{\text{\AA}}
\DeclareSIUnit\au{\text{at.\,u.}}
\usepackage{amsmath,amssymb}

\begin{document}

\title{Statistical inference of collision frequencies from x-ray Thomson scattering spectra}

\author{Thomas W. Hentschel}
\affiliation{School of Applied \& Engineering Physics, Cornell University, Ithaca NY, 14850 USA \looseness=-1}
\author{Alina Kononov}
\affiliation{Center for Computing Research, Sandia National Laboratories, Albuquerque NM, 87123 USA \looseness=-1}
\author{Andrew D. Baczewski}
\affiliation{Center for Computing Research, Sandia National Laboratories, Albuquerque NM, 87123 USA \looseness=-1}
\author{Stephanie B. Hansen}
\affiliation{Pulsed Power Sciences Center, Sandia National Laboratories, Albuquerque NM, 87123 USA \looseness=-1}

\begin{abstract}
Thomson scattering spectra measure the response of plasma particles to incident radiation. In warm dense matter, which is opaque to visible light, x-ray Thomson scattering (XRTS) enables a detailed probe of the electron distribution and has been used as a diagnostic for electron temperature, density, and plasma ionization. In this work, we examine the sensitivities of inelastic XRTS signatures to modeling details including the dynamic collision frequency and the electronic density of states. Applying verified Monte Carlo inversion methods to dynamic structure factors obtained from time-dependent density functional theory, we assess the utility of XRTS signals as a way to inform the dynamic collision frequency, especially its direct-current (DC) limit, which is directly related to the electrical conductivity.
\end{abstract}

\maketitle

\section{Introduction}

The behavior of free electrons often dominates important response properties of high-energy-density (HED) matter, ranging from conductivity to alpha-particle stopping power in inertial confinement fusion.
Models of such response properties provide critical input for hydrodynamic simulations of HED experiments\cite{haines_charged_2024} and enable interpretation of experimental diagnostics like x-ray Thomson scattering (XRTS)\cite{glenzer_observations_2007}.

However, the dynamic response of free electrons evades accurate and efficient modeling in the warm dense regime, where partial ionization, degeneracy, and density effects distort electronic structure; strong coupling and ion correlations become important; and electron-ion scattering cannot be neglected.
For example, the random phase approximation (RPA), which treats ions as a uniform background charge and electrons as a uniform electron gas, overestimates both the energy shift and magnitude of the plasmon peak in XRTS spectra \cite{sperling2015free,hentschel:2023}.
The Mermin approximation \cite{mermin1970} to the dielectric function goes beyond the RPA by including the effects of a dynamic (frequency-dependent) electron-ion collision rate, which can shift and broaden XRTS signatures \cite{hentschel:2023}.

Electron-ion collision rates are also a key factor in other response and transport properties, such as the direct-current (DC) electrical conductivity. Collisions have been extensively studied in the classical plasma regime \cite{spitzer}, warm dense matter \cite{Mazvet2005,JohnsonKG}, and liquid metals \cite{ziman1961}. In the warm dense regime, electron-ion collision rates can be calculated using average-atom models, where the rates depend on electron-ion impact cross sections, ion structure factors, and the free-electron density of states (DOS) \cite{RinkerPRA88, Burrill, PainWetta2020}. These zero-frequency DC collision rates can be extended to the dynamic (frequency-dependent) regime using the Born \cite{Thiele2008,Reinholz2000} or Lenard-Balescu \cite{faussurier2016electron} approaches and used as input to the Mermin approximation to produce dynamic structure factors (DSFs), which are closely related to XRTS spectra \cite{johnson2012thomson,Souza2014,hentschel:2023}.

State-of-the-art first-principles methods like time-dependent density functional theory (TDDFT) can predict both XRTS spectra \cite{baczewski_x-ray_2016,baczewski_predictions_2021} and dynamic conductivities \cite{andrade_negative_2018} without relying on an electron-ion collision rate model.
These methods carry high computational costs and depend on other fundamental modeling choices such as exchange-correlation functionals and pseudpotentials.
Recent work proposed indirectly extracting dynamic collision frequencies from first-principles simulations by solving an optimization problem to match the Mermin dielectric ansatz to optical properties predicted by density functional theory \cite{schorner2023xray}.

Experimentally validating collisional models --- along with the various choices and approximations involved --- is difficult because experiments do not access electron-ion collision rates directly. 
Validating collision frequencies is thus typically done through a forward-modeling approach, inputting a modeled dynamic collision frequency into the Mermin dielectric function, computing observable quantities like XRTS spectra, and comparing the results to experiments \cite{sperling2015free} or first-principles simulations \cite{witte2017warm,hentschel:2023}. Alternatively, within a sufficiently constrained modeling framework (e.g. the Mermin dielectric ansatz and a known electronic density of states), measured XRTS data can be inverted to estimate a dynamic collision frequency \cite{sperling_electrical_2017}. 
Both of these approaches have limitations: a constructed or extracted collision frequency that adequately fits an experimental scattering signal is not guaranteed to be a unique solution, and the limited range of options in a direct calculation of collision frequencies is not guaranteed to generate a collision frequency that can adequately fit a given scattering signal.

In this work, we address these limitations by using Bayesian statistics to infer dynamic collision frequencies that are based on a highly flexible parameterized model. This method allows us to assess the sensitivities of the extracted collision rates to the frequency range and uncertainties of the observable XRTS signals. It also provides rigorous uncertainty bounds as a function of frequency that can help inform uncertainties in transport coefficients extracted from XRTS data. We describe the parameterized collision-frequency model and Bayesian inference method in Section \ref{sec:methods}. We apply the method to dynamic structure factors generated by TDDFT simulations for warm dense aluminum in Section \ref{sec:results}, and we summarize our findings in Section \ref{sec:conclusions}. 

\section{Methods}
\label{sec:methods}

\subsection{Collision frequency model}
In order to numerically infer a collision rate function from XRTS data, we choose a functional form that is flexible enough to describe a large class of potential collision rate models while also satisfying physical constraints.
Throughout the article, we assume that the collision frequency obeys the Kramers-Kronig relations, so the imaginary part of the collision frequency is completely determined by the real part. 
We therefore parameterize the real part only.
Since the real part of a function obeying Kramers-Kronig relations must be symmetric, we further impose $\nu(-\omega) = \nu(\omega)$ and consider only $\omega>0$.

The first term in our model is inspired by the Born collision frequency, which resembles a Lorentzian-like function centered at $\omega =0$ with height $\nu_0$, width controlled by $b_0$, and a $3/2$ power decay:
\begin{equation}
    \nu_b(\omega;\nu_0) = \frac{\nu_0}{1 + (\omega / b_0)^{3/2}}.
    \label{eq:collision-born}
\end{equation}
This term represents free electrons scattering weakly from ions.
We constrain the width $b_0$ by enforcing a sum rule 
(see Appendix~\ref{app:collision-sum-rule}).
Thus, $\nu_0$ is the only free parameter in $\nu_b$.

At higher frequencies, other collisional processes may occur that are not accounted for by a simple Born picture of electron-ion collisions, like collisions involving non-ideal free electrons and inelastic scattering processes \cite{hentschel:2023}.
To model the influence of these mechanisms, we include an additional term where a modified Born-like function ``activates'' at a particular frequency $\omega_a$:
\begin{equation}
    \nu_a(\omega; \nu_1, \omega_a, \alpha, p) = \frac{\nu_1}{1+e^{-(\omega - \omega_a)/\alpha} + (\omega/\omega_a)^p}.
    \label{eq:collision-nonideal}
\end{equation}
The logistic component produces this onset behavior, with $\alpha$ governing its steepness.
Finally, we allow for more flexibility in the secondary Born-like decay by replacing the $3/2$ exponent in the denominator by another parameter $p$.
All of this will allow our model to describe non-ideal and inelastic collisions. 

Combining Eqs.~\eqref{eq:collision-born} and \eqref{eq:collision-nonideal}, the final model has five adjustable parameters $\Theta = (\nu_0, \nu_1, \omega_a, \alpha, p)$:
\begin{equation}
   \mathrm{Re}[\nu_m(\omega;\Theta)] = \nu_b(\omega; \nu_0) + \nu_a(\omega; \nu_1, \omega_a, \alpha, p).
    \label{eq:collision-model}
\end{equation}
We note that in the DC limit, this ansatz reduces to
\begin{equation}
    \mathrm{Re}[\nu_m(\omega=0;\Theta)] =
    \nu_0 + \frac{\nu_1}{1 + e^{\omega_a/\alpha}} \approx \nu_0,
\end{equation}
where the approximate form holds for $\omega_a \gg \alpha$.
Although the model form does not explicitly enforce smoothness at $\omega=0$, we also have $\partial_\omega \mathrm{Re}[\nu_m] |_{\omega=0} \approx 0$ in the physically relevant $\omega_a \gg \alpha$ and $p>1$ parameter regime.
Meanwhile, the limiting behavior at high frequencies ($\omega\gg b_0$ and $\omega \gg \omega_a$) is $\mathrm{Re}[\nu_m]\sim \omega^{-p_1}$ with $p_1 = \min(p,3/2)$.

The parameters in the model are restricted to positive quantities.
This constraint is physically motivated for $\nu_0$ and $\nu_1$, since the real part of the collision frequency must be non-negative.
Additionally, $p > 0$ ensures that $\nu(\omega)$ decays at high frequencies and produces a well-defined imaginary part $\mathrm{Im}[\nu_m]$ through the Kramers-Kronig relations.
We also restrict $p < 3/2$ so that the high-frequency limit ($\omega\gg b_0$ and $\omega \gg \omega_a$) is governed by $\nu_a$ for non-zero $\nu_1$.
Finally, $\alpha$ and $\omega_a$ are positive so that $\nu_a$ captures physics beyond the Born approximation at positive $\omega$.

\begin{figure}
    \includegraphics[width=\columnwidth]{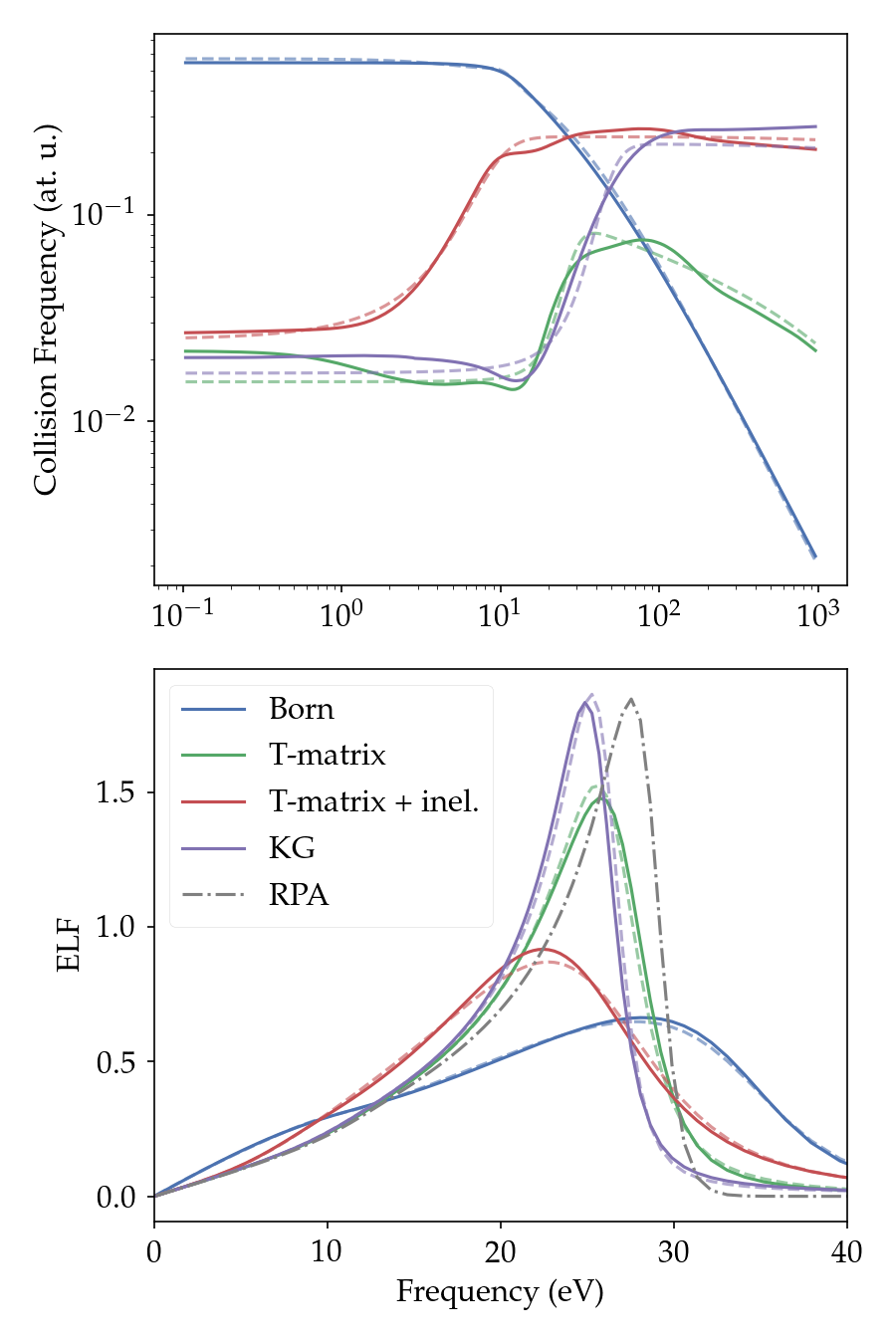}
    \caption{(Top) Direct fits of our model for the real part of the collision frequency (dashed curves) to predictions from different theories considered by Ref.~\onlinecite{hentschel:2023} for solid-density aluminum at a temperature of \SI{1}{\electronvolt} and momentum transfer of \SI{1.55}{\per\angstrom} (solid curves).
    The blue solid curve is based on a Born cross section, the purple curve represents an inversion of the dynamic conductivity from a Kubo-Greenwood treatment, the green and red solid curves are based on T-matrix cross-sections and nontrivial ion structure factors, and the red curve includes additional inelastic processes \cite{hentschel:2023}.   
    The dashed curves represent corresponding direct nonlinear least-squares fits of the model described by Eqs.\ \eqref{eq:collision-born}--\eqref{eq:collision-model}.
    (Bottom) Corresponding ELFs computed using the Mermin dielectric function (Eq.~\eqref{eq:mermin-dielectric}).
    The gray curve is calculated using the RPA dielectric function, which is equivalent to the Mermin dielectric with a collision frequency of zero.
    }
    \label{fig:model-flexibility}
\end{figure}

In Fig.~\ref{fig:model-flexibility}, we demonstrate the flexibility of our collision frequency model by directly fitting it to the collision frequency theories discussed in Ref.~\onlinecite{hentschel:2023} and summarized in the the figure caption.
With appropriately optimized parameters $\Theta$, the model $\nu_m(\omega;\Theta)$ (dashed curves) successfully captures the main features of the the various collision frequency theories (solid curves). 

The collision frequency informs dynamic structure factors and XRTS spectra through the Mermin dielectric function \cite{mermin1970},
\begin{equation}
    \epsilon_M(q, \omega; \nu) = 1 + \frac{(\omega+i\nu)[\epsilon_\mathrm{RPA}(q, \omega+i\nu) - 1]}{\omega + i\nu \frac{\epsilon_\mathrm{RPA}(q, \omega+i\nu) - 1}{\epsilon_\mathrm{RPA}(q,0)-1}}  \;,
    \label{eq:mermin-dielectric}
\end{equation}
where $q$ and $\omega$ are the momentum and energy transferred by the scattered x-rays, $\epsilon_\mathrm{RPA}$ is the RPA dielectric function, and $\nu(\omega)$ is the dynamic collision frequency.
Then, the DSF is given by
\begin{equation}
    S_M(q, \omega; \nu) = -\frac{1}{1 - e^{-\omega/k_B T}} \, \frac{q^2}{4 \pi Z_0 n_i} \,\mathrm{ELF}_M(q,\omega; \nu),
    \label{eq:dsf}
\end{equation}
where
\begin{equation}
    \mathrm{ELF}_M(q,\omega; \nu) = \mathrm{Im}\left[ -\frac{1}{\epsilon_M(q, \omega; \nu)}\right]
    \label{eq:elf}
\end{equation}
is the energy loss function (ELF) corresponding to the Mermin dielectric function.
For $\nu(\omega)=0$, $\epsilon_M$ reduces to $\epsilon_\mathrm{RPA}$.

Different choices for the collision frequency $\nu(\omega)$ will influence the ELF and DSF derived from the Mermin dielectric function, as illustrated in Fig.~\ref{fig:model-flexibility}. In general, we find that the ELF is locally broadened by the real part of the dynamic collision frequency and shifted by the imaginary part.
The real part of the Born collision frequency decreases monotonically, leading to an imaginary collision frequency that is everywhere positive. Thus the peak of the corresponding ELF is blueshifted to higher energies relative to the collisionless RPA. In contrast, the imaginary part of collision frequencies with logistic contributions can have negative values, which induce redshifts in the plasmon peak. 
Our parameterized model collision frequency function $\nu_m$ can reproduce each of these potential modifications to the RPA. We find that the minor deviations from the original collision frequencies lead to minor differences in the resulting ELFs. 

\subsection{Bayesian inference of collision frequencies}

Our simple, yet flexible, five-parameter model $\nu_m(\omega;\Theta)$ enables efficient inversion to constrain collision frequencies using first-principles or experimental reference data for an electronic response function (e.g., XRTS spectrum).
Since the prefactor to the ELF in Eq.\ \eqref{eq:dsf} does not depend on the collision frequency and we will focus on the $\omega > k_BT$ regime, we use the ELF (rather than the DSF) as our objective function throughout this work. 

Inverting the Mermin ELF to infer collision frequencies is a poorly-constrained optimization problem because, as we will see, there are often many distinct choices for $\nu(\omega)$ that produce an ELF consistent with the reference data.
Any uncertainties or noise in the reference data would further compound the difficulty of determining an optimal parameter set $\Theta$.
Instead, we use a Bayesian approach to identify parameter distributions that generate Mermin ELFs close to the reference data set.

Within the Bayesian framework, the posterior $p(\Theta | y(\omega))$ describes the distribution of parameters $\Theta=(\nu_0, \nu_1, \omega_a, \alpha, p)$ given the reference ELF data $y = (y_1, \ldots, y_N)$ over the frequency grid $\omega = (\omega_1, \ldots, \omega_N)$.
Bayes' theorem relates the posterior distribution to the prior distribution and likelihood function: 
\begin{equation}
    p(\Theta | y(\omega)) \propto p(y(\omega)| \Theta) \, p(\Theta).
    \label{eq:bayes}
\end{equation}
We use a multivariate uniform distribution for the prior distribution $p(\Theta)$ with the following parameter ranges: $\nu_0 \in [0,5]$, $\nu_1 \in [0,5]$, $\omega_a \in [0, 40]$, $\alpha \in [10^{-3}, 10]$, and $p \in [0, 1.5]$, where all of the parameters are in atomic units except $p$, which is unitless.
The lower bounds for these intervals restrict the parameters to nonnegative values to satisfy physical constraints, with $\alpha\geq 10^{-3}$ further accounting for the finite frequency resolution.
Meanwhile, the upper bounds are sufficiently large to generously encompass most parameter values that produce ELFs consistent with the reference dataset.

For the likelihood function, we choose a squared-exponential form that is closely related to the loss function used in least-squares optimization problems:
\begin{equation}
    \label{eq:likelihood-sqexp}
    p(y(\omega) | \Theta) \propto \exp\left(-\frac{\sum_i | r_i(\Theta) |^2}{2\sigma^2}\right),
\end{equation}
where 
\begin{equation}
    r_i(\Theta) = y_i - \mathrm{ELF}_M(\omega_i; \nu_m(\omega_i,\Theta))
\end{equation}
is the absolute residual between the reference ELF data $y_i$ and the corresponding Mermin ELF forward model prediction for a given $q$ (see Eq.~\eqref{eq:elf}).
The standard deviation $\sigma$ determines how closely the Mermin ELFs corresponding to the probable $\Theta$ samples match the reference data.
For a particular piece of reference data, one might set $\sigma$ according to the specific noise level; here, we used $\sigma = 0.1$, a value around 10\% of the reference ELF peak, to represent a typical noise level for XRTS data from both experiments (which are challenged by small scattering cross sections, limited probe intensities, and significant background) and first-principles simulations (which can vary with atomic configuration; see Appendix \ref{app:tddft}).

When the ELF intensity spans multiple orders of magnitude, the residual in Eq.~\eqref{eq:likelihood-sqexp} will naturally emphasize agreement with larger reference ELF values $y_i$.
This ``absolute'' residual is appropriate for noisy data where we wish to find a good fit to the the ELF near the plasmon peak and where signals below the noise level are obscured by the noise.

We also wish to test the inversion method in an ideal case, where we wish to treat all the data as equally important.
For that case, we use a ``relative'' residual 
\begin{equation}
   r_i^\mathrm{rel}(\Theta) = \frac{y_i - \mathrm{ELF}_M(\omega_i; \nu_m(\omega_i,\Theta))}{y_i}
   \label{eq:relative-residual}
\end{equation}
which roughly equalizes the weight of the residual regardless of the magnitude of the data $y_i$. 

Many of the features of the posterior distribution we are interested in --- like the mean to determine the average set of parameters or the standard deviation to estimate parameter uncertainties --- involve computing intractable multidimensional integrals. Instead of evaluating these integrals directly, we use Markov chain Monte Carlo (MCMC) to draw random samples from the posterior, which we can use to approximate these quantities.
We use the MCMC ensemble sampler implemented in the Python library \textsc{emcee}~\cite{foremanmackey2013emcee,goodman2010ensemble} for this purpose.
We also use the mean autocorrelation time to evaluate the sampling error and the robustness of the analysis, as recommended by Ref.~\onlinecite{foremanmackey2013emcee}.

\section{Results and Discussion}
\label{sec:results}

\begin{figure}
    \centering
    \includegraphics[width=\columnwidth]{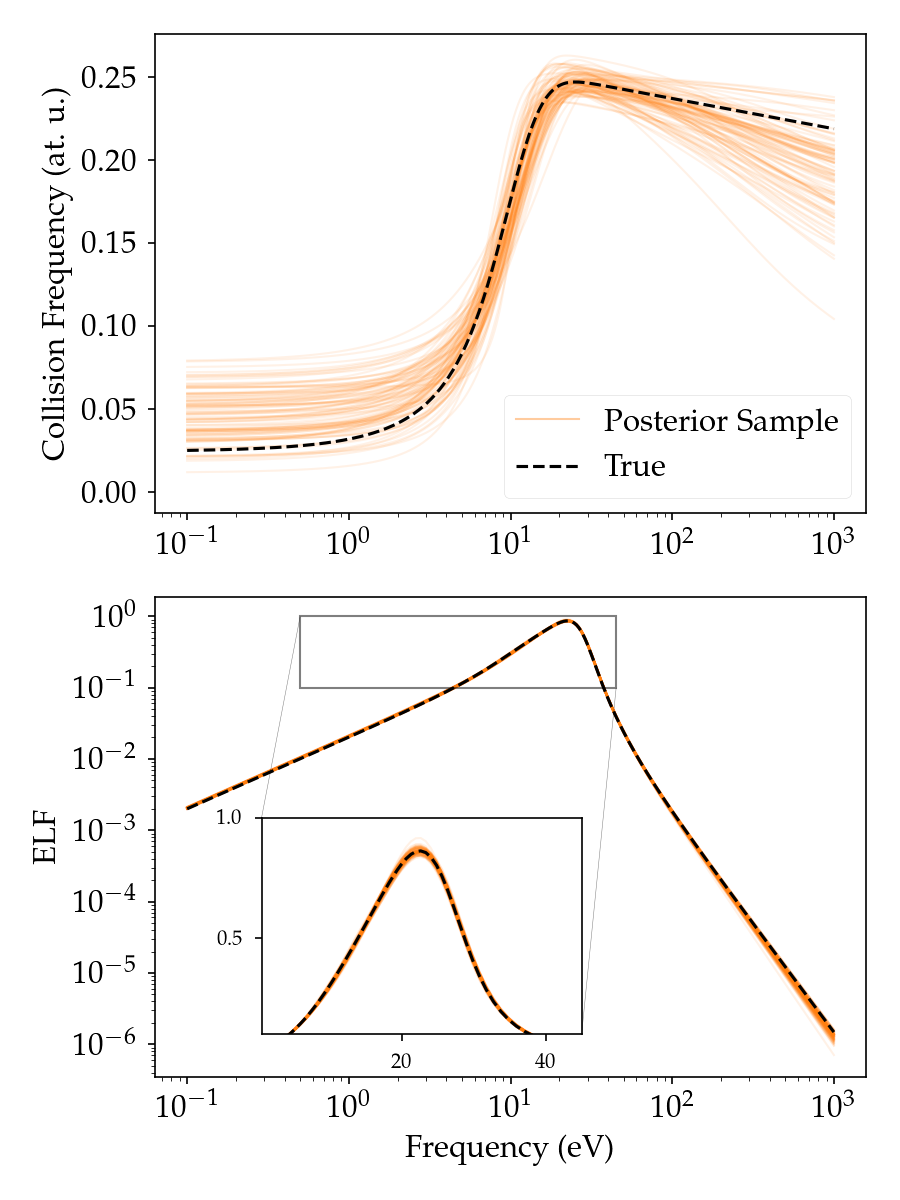}
    \caption{
    Inference results for the real part of the dynamic collision frequency (top) and corresponding ELF (bottom) for an idealized scenario where the reference ELF $y_0(\omega)$ (dashed black) was generated using the model $\nu_m(\omega;\Theta_0)$.
    The parameters $\Theta_0$ were obtained from a direct fit to a collision frequency theory based on T-matrix cross-sections with inelastic processes\cite{hentschel:2023} (see red curves in Fig.~\ref{fig:model-flexibility}).
    Orange curves show the inferred $\nu_m(\omega;\Theta)$ and corresponding Mermin ELFs for $\Theta$ parameters sampled from the posterior distribution $p(\Theta | y_0(\omega))$ using the relative residual in the likelihood function over the entire frequency range.
    The inset highlights behavior near the ELF peak on a linear scale. 
    The ELFs correspond to a wavenumber of $q = \SI{1.55}{\per\angstrom}$ for solid-density aluminum at a temperature of \SI{1}{\electronvolt}, and the Mermin ELFs use an ideal DOS with a chemical potential of \SI{11.6}{\electronvolt}.
    }
    \label{fig:ideal-inference}
\end{figure}

First, we assess inherent uncertainties in inferring collision frequencies by considering an idealized situation with noise-free ELF data that can be exactly described by the collision frequency model used in the inference (Eq.~\eqref{eq:collision-model}).
That is, we use a reference collision frequency $\nu_0 = \nu_m(\omega;\Theta_0)$ from one of the direct fits in Fig.~\ref{fig:model-flexibility} to generate the reference ELF data $y_0(\omega)=\mathrm{ELF}_M(q,\omega;\nu_0)$ (dashed black curves in Fig.~\ref{fig:ideal-inference}) \footnote{The parameters determining $\nu_m(\omega;\Theta_0)$ and the corresponding reference ELF are $\Theta_0 = (\num{1e-6}, \num{5.1e-1}, \num{3.6e-1}, \num{1.2e-1}, \num{6.3e-2})$.}.
Using MCMC with a relative residual in the likelihood function, we then sample parameters $\Theta$ from the posterior distribution $p(\Theta | y_0(\omega))$ to obtain other model dynamic collision frequencies that yield ELFs consistent with the reference ELF (orange curves in Fig.~\ref{fig:ideal-inference}).

Even in this ideal case, we find that a fairly large range of collision frequencies can give excellent agreement with the reference ELF. 
Moreover, the spread among inferred collision frequencies in Fig.~\ref{fig:ideal-inference} represents a lower bound for uncertainties in more realistic scenarios, where noise in the reference data could expand what counts as an acceptable fit and  behavior beyond the Mermin approximation could introduce additional ambiguities in extracting representative collision frequencies.
Notably, the DC limit of the collision frequency is particularly ill-constrained --- here, we obtain excellent fits to the objective ELF with factor-of-three variations in the DC limit.

\begin{figure}
    \centering
    \includegraphics[width=\columnwidth]{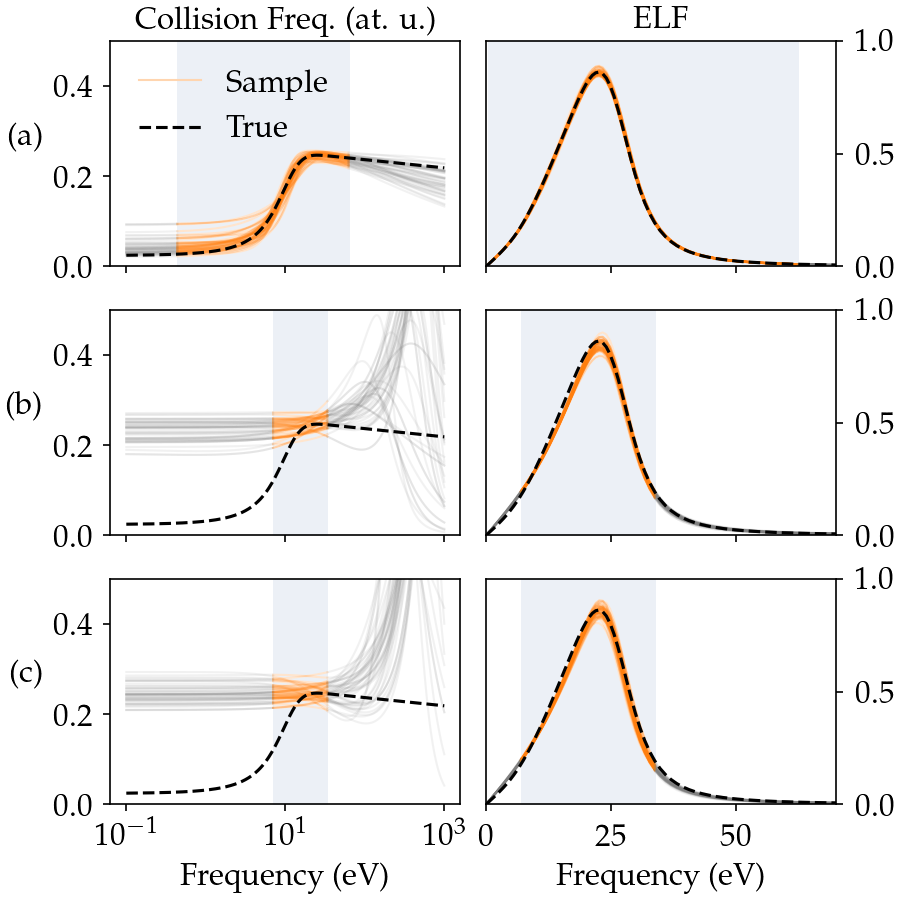}
    \caption{
    Effects of modifying the inference framework for applicability to realistic reference data.
    For the same idealized scenario as Fig.~\ref{fig:ideal-inference},
    row (a) uses the relative residual in the likelihood function and considers data within 99\% of the reference ELF peak;
    row (b) uses the relative residual and considers data within 80\% of the peak;
    and row (c) uses the absolute residual and considers data within 80\% of the peak. 
    The shaded background and orange lines represent the frequency range of ELF data considered in each case, while the gray portion of the collision frequency and ELF curves indicate frequencies outside of this range.}
    \label{fig:ideal-inference-variations}
\end{figure}

To assess the reliability of our inversion method for less ideal cases, we explore three systematic deviations from the idealized framework, illustrated in Fig.~\ref{fig:ideal-inference-variations}. 
We first restrict the frequency range of our fit to the reference data.
This modification is important because the Mermin model gives only the free-free contribution to the total ELF, while experimental spectra can also contain contributions from quasielastic scattering at low frequencies and bound-free transitions at high frequencies.
While first-principles TDDFT calculations can isolate free-free contributions through appropriate pseudopotentials \cite{baczewski_x-ray_2016}, convergence difficulties and numerical sensitivities can cause significant uncertainties at both frequency extremes.

We explore the frequency restriction in two steps, retaining the relative residual suitable for our noise-free reference data.
First, we fit only to data within 99\% of the reference ELF maximum in Fig.~\ref{fig:ideal-inference-variations}a.
This choice leads to collision frequencies very similar to those obtained for the full frequency range (Fig.~\ref{fig:ideal-inference}).
Next, we fit only to data within 80\% of the peak in Fig.~\ref{fig:ideal-inference-variations}b.
This more extreme restriction significantly changes the inferred DC limit --- shifting DC values up by about a factor of five ---  and leaves the high-frequency regime largely unconstrained.

Finally, retaining the more restricted frequency range, we replace the relative residual with the absolute residual in Fig.~\ref{fig:ideal-inference-variations}c.
This adjustment produces collision frequencies very similar to those using the relative residual over the same range, with a small shift in the mean DC value and a modest reduction of variances outside of the restricted frequency range.

Overall, we find that imposing realistic frequency cutoffs dramatically degrades the ability to recover the reference collision frequency away from the plasmon peak even though the MCMC samples reproduce the reference ELF very closely. Further, while the highly range-restricted cases roughly capture local values of the reference collision frequency near the plasmon peak, their variations about the DC limit are not representative of the full range of collision frequencies that can reproduce the reference ELF. 

We now apply the Bayesian approach used in Fig.~\ref{fig:ideal-inference-variations}c to infer collision frequencies consistent with first-principles ELFs computed from TDDFT for solid-density aluminum at a temperature of \SI{1}{\electronvolt}.
The TDDFT simulations compute the ELF from the real-time electronic response to a perturbation (see Appendix \ref{app:tddft} for details) without relying upon the Mermin dielectric ansatz.
Although TDDFT excels at capturing electron-ion interactions --- including anisotropic density effects beyond typical average-atom treatments --- to our knowledge this method cannot directly access electron-ion collision frequencies.
Nonetheless, indirectly inferring collision rates from TDDFT response functions could allow more detailed benchmarking of Mermin-based models or inform a surrogate model to efficiently predict response properties with first-principles accuracy over a wider range of conditions.

\begin{figure}
    \centering
    \includegraphics[width=\columnwidth]{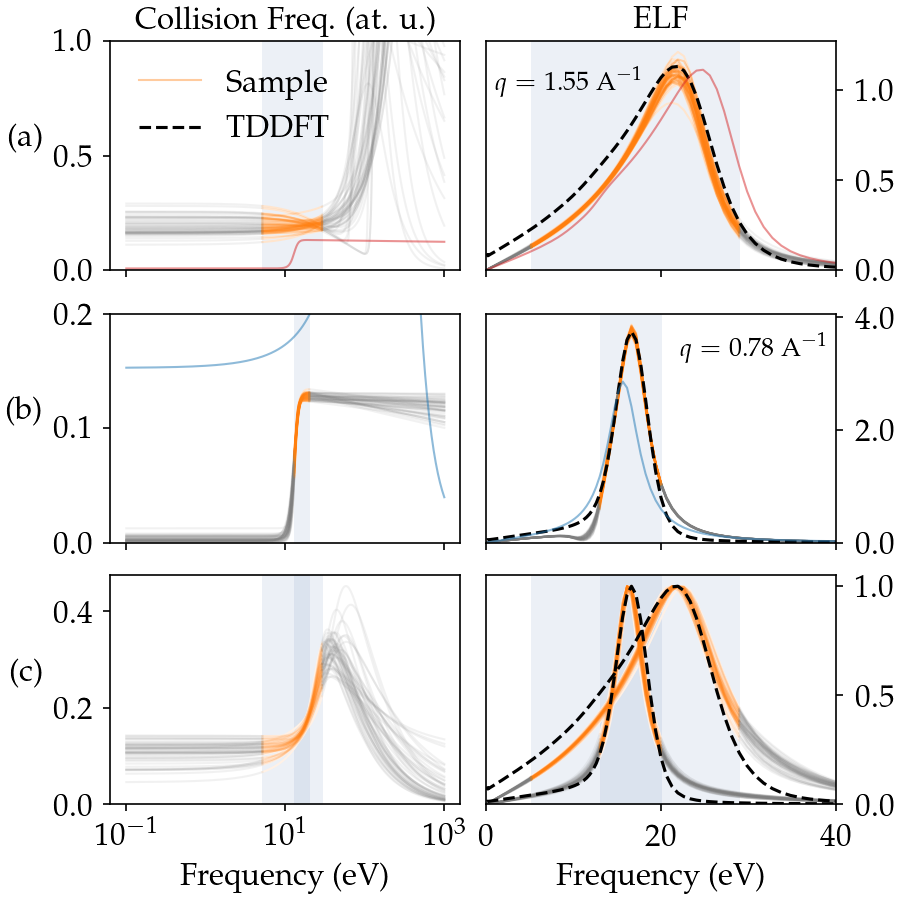}
    \caption{
    Inferred collision frequencies and corresponding Mermin ELFs when using first-principles predictions from TDDFT (dashed black) as reference data and applying the same methods used in Fig.~\ref{fig:ideal-inference-variations}c to emphasize agreement near the plasmon peak.
    Orange curves show inference results when considering TDDFT data for (a) $q = \SI{1.55}{\per\angstrom}$, (b) $q = \SI{0.78}{\per\angstrom}$, and (c) both wavenumbers simultaneously. 
    The red curve in row (a) uses a typical collision frequency inferred in row (b) for $q = \SI{0.78}{\per\angstrom}$ to evaluate the Mermin ELF at $q = \SI{1.55}{\per\angstrom}$, and vice versa for the blue curve in row (b).
    Background shading indicates the frequency range considered for each ELF, and the heights of both ELFs have been normalized to 1 in row (c).
    }
    \label{fig:tddft-inference}
\end{figure}

Figure~\ref{fig:tddft-inference}a shows that the collision frequencies inferred from TDDFT data for $q = \SI{1.55}{\per\angstrom}$ qualitatively resemble the results of the idealized scenario for the same wavenumber. However, the corresponding ELFs tend to underestimate the TDDFT ELF for $\omega \lesssim \SI{18}{\electronvolt}$, resulting in a somewhat narrower plasmon peak.
This discrepancy suggests that the TDDFT predictions may contain physics beyond the Mermin ansatz.

Inference results for a lower wavenumber of \SI{0.78}{\per\angstrom} exhibit several notable differences from behavior at \mbox{$q=\SI{1.55}{\per\angstrom}$} (see Fig.~\ref{fig:tddft-inference}b).
First, the inferred collision frequencies now capture the TDDFT ELF extremely closely within the frequency range used to evaluate the residual, though deviations do appear in the low- and high-frequency tails beyond the range included in the analysis.
The TDDFT data constrains the collision frequency much more tightly in this lower-$q$ case, with considerably less spread among the $\nu_m(\omega;\Theta)$ samples both within and outside of the frequency range considered.

Interestingly, the inferred collision frequencies for the two wavenumbers differ both in terms of their qualitative shape and their quantitative values.
Therefore, a collision frequency inferred from a response function at one $q$ value cannot necessarily predict properties at a different $q$ value.
Indeed, using the inference results corresponding to $q = \SI{0.78}{\per\angstrom}$ or $\SI{1.55}{\per\angstrom}$ to evaluate the Mermin ELF at the other wavenumber fails to adequately capture the reference TDDFT data (see red and blue curves in Fig.~\ref{fig:tddft-inference}a and b).
That is, with a realistically restricted frequency range for the inversion, a single scattering angle may not provide enough information to determine a universally applicable collision frequency.

To explore whether some of the information lost by restricting the frequency range can be recovered by providing data at more than one scattering angle, we perform a simultaneous fit to both wavenumbers in Fig.~\ref{fig:tddft-inference}c.
This analysis produces collision frequencies that bear little resemblance to the inference results for either wavenumber considered individually --- but which, intriguingly, more closely resemble the non-Born curves of Fig.~\ref{fig:model-flexibility}. 
Given the lack of overlap between the single-angle collision frequency distributions (see Fig.~\ref{fig:tddft-inference}a and b), the two-angle inference tends to sample intermediate values to balance agreement with both TDDFT ELFs.
The corresponding Mermin ELFs still capture the TDDFT data at both angles quite well, despite a slight redshift of the $q = \SI{0.78}{\per\angstrom}$ plasmon peak and a more gradual decay of the $q = \SI{1.55}{\per\angstrom}$ high-energy tail compared to the single-angle results.

We note that both the single-angle and multi-angle inversions give ELFs that underpredict the low-frequency values of the $q=\SI{1.55}{\per\angstrom}$ ELF from TDDFT, suggesting that the forward model may be incomplete. A known deficiency of the RPA dielectric function underlying the Mermin model (see Eq.~\eqref{eq:mermin-dielectric}) is that it assumes a uniform electron gas, neglecting the influence of ions on the electronic structure.
To overcome this deficiency, Ref.~\onlinecite{hentschel:2023} proposed modifying $\epsilon_\mathrm{RPA}$ with the quantum density of states (DOS) obtained from the Kohn-Sham continuum orbitals from an average-atom calculation. The quantum DOS is roughly 3/2 larger than the ideal DOS, requires a smaller chemical potential to enforce charge neutrality in the ion sphere, and better matches the DOS from multi-center DFT calculations\cite{hentschel:2023}.  
To test the effect of the DOS on the collision inference, we adjust the chemical potential and modify the $\epsilon_\mathrm{RPA}$ in Eq.~\eqref{eq:mermin-dielectric} with a factor that recovers the quantum DOS. 

\begin{figure}
    \centering
    \includegraphics[width=\columnwidth]{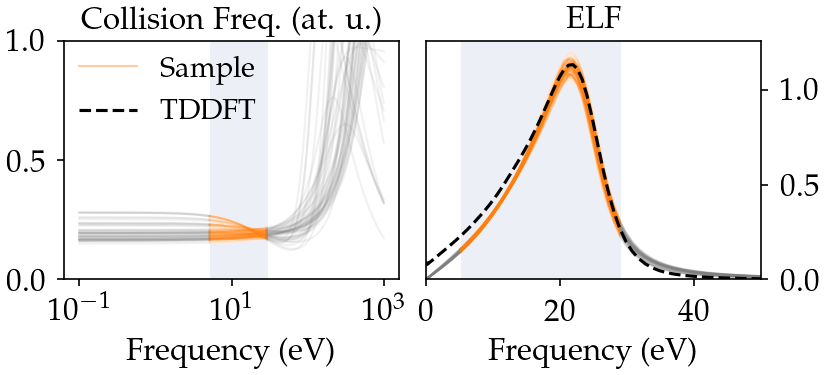}
    \caption{
    Inference results for the same case as Fig.~\ref{fig:tddft-inference}a, but now using a modified forward Mermin model that incorporates a nonideal DOS computed from an average-atom model.
    The likelihood function uses the absolute residual and only data within 80\% of the peak value is considered.
    }
    \label{fig:tddft-inference-al-nonideal}
\end{figure}

Repeating the inference shown in Fig.~\ref{fig:tddft-inference}a with the modified forward model, we find improved agreement with the TDDFT ELF (see Fig.~\ref{fig:tddft-inference-al-nonideal}).
However, using the quantum DOS has only a minor effect on the inferred collision frequencies for this case.
The electronic structure treatment underlying the Mermin dielectric function may become more important for systems with stronger departures from free electron gas behavior.

\begin{figure}
    \centering
    \includegraphics[width=\columnwidth]{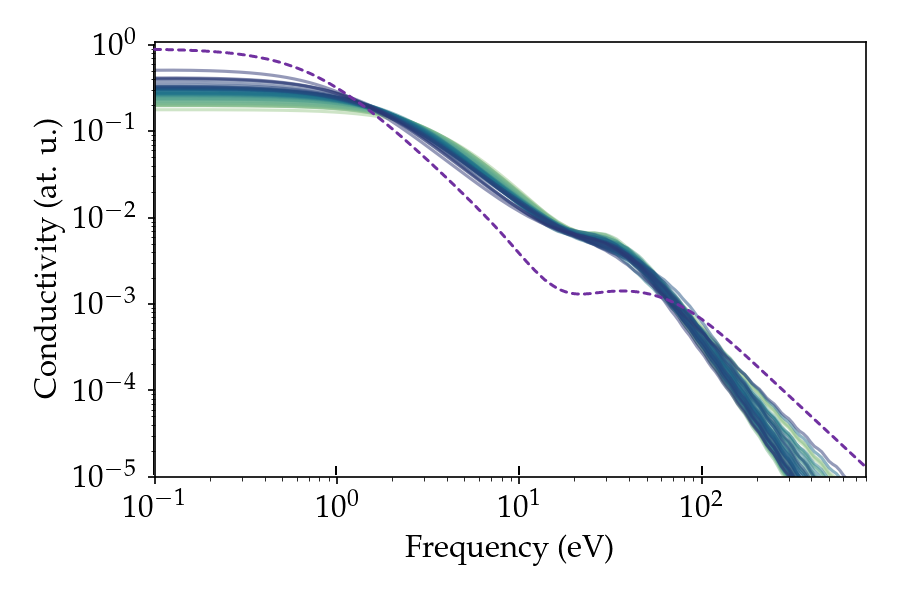}
    \caption{
    Inferred dynamic conductivities, in atomic units, computed using the Mermin dielectric function according to Eq.~\eqref{eq:conductivity}.
    The collision frequencies were inferred from TDDFT ELF data for two wavenumbers considered simultaneously (see Fig.~\ref{fig:tddft-inference}c). The dashed purple line is a direct calculation of the dynamic conductivity from an average-atom model using the Kubo-Greenwood formalism, as previously published in Ref.~\onlinecite{hansen_nlte_report}. 
    }
    \label{fig:conductivity}
\end{figure}

Finally, we assess the prospects of using scattering data to constrain conductivity through the DC limit of the inferred electron-ion collision frequencies.
The results of Figs.~\ref{fig:ideal-inference-variations} and \ref{fig:tddft-inference} already indicate that the scattering angle and frequency range of useful data may sensitively influence an inferred DC conductivity.
More concretely, the dynamic conductivity depends on the $q\rightarrow 0$ limit of the dielectric function:
\begin{equation}
    \sigma(\omega) = \frac{\omega}{4\pi i} (\epsilon(0,\omega) - 1).
    \label{eq:conductivity}
\end{equation}
Figure \ref{fig:conductivity} evaluates Eq.~\eqref{eq:conductivity} using the Mermin dielectric function with $\nu_m(\omega;\Theta)$ samples from the two-angle TDDFT inference of Fig.~\ref{fig:tddft-inference}c.
Notably, this curve exhibits non-Drude behavior above about 10 eV, similar to that seen in previous DFT-MD calculations \cite{witte2017warm} and even in the Kubo-Greenwood conductivity from the average-atom model (shown in the dashed purple line).

However, the spread among predictions for the DC limit of the conductivity in Fig.~\ref{fig:conductivity} spans about a factor of $2$.
The collision frequency samples can be used to estimate a mean $\overline{\sigma}_\mathrm{DC}$ and standard deviation  $\Delta\sigma_\mathrm{DC}$ for the inferred DC conductivity:
\begin{align}
    \overline{\sigma}_\mathrm{DC} 
      &= \int \sigma_\mathrm{DC}(\Theta) \, p(\Theta | y(\omega)) \, d\Theta \nonumber\\
      &\approx \frac{1}{M} \sum_{m=1}^M \sigma_\mathrm{DC}(\Theta_m),
      \label{eq:sigma_mean}
\end{align}
\begin{align}
    \Delta\sigma_\mathrm{DC}^2
      &= \int (\sigma_\mathrm{DC}(\Theta) - \overline{\sigma}_\mathrm{DC})^2 \, p(\Theta | y(\omega)) \, d\Theta \nonumber\\
      &\approx \frac{1}{M} \sum_{m=1}^M (\sigma_\mathrm{DC}(\Theta_m) - \overline{\sigma}_\mathrm{DC})^2,
      \label{eq:sigma_std}
\end{align}
where 
$\sigma_\mathrm{DC}(\Theta) = n_e/\nu_m(0; \Theta)$
is the DC conductivity as a function of the collision frequency parameters $\Theta$, \mbox{$n_e = 0.027$} is the electron density in atomic units assuming 3 free electrons per ion in solid-density aluminum, $p(\Theta | y(\omega))$ is the posterior distribution function, and $M$ is the number of MCMC samples $\Theta_m$ drawn from $p(\Theta | y(\omega))$.

In the idealized scenario of Fig.~\ref{fig:ideal-inference},
the computed DC conductivity has a mean of 0.64 atomic units (\SI{3.0E+6}{\siemens\per\meter}) and a standard deviation of 0.31 atomic units (\SI{1.4E+06}{\siemens\per\meter}).
The roughly 50\% spread about the mean even when performing the inference on an ELF within the model space and over the full frequency range already illustrates the difficulty of inferring a precise DC conductivity from $q>0$ response properties.

Table \ref{tab:DCconductivity} lists the means and standard deviations of the DC conductivities inferred from TDDFT data via the electron-ion collision frequencies shown in Fig.~\ref{fig:tddft-inference}. 
For the $q = \SI{0.78}{\per\angstrom}$ case corresponding to Fig.~\ref{fig:tddft-inference}b, the lower limit of the DC collision frequency and thus the upper limit of the DC conductivity are effectively unconstrained, generating a very large $\Delta\sigma_\mathrm{DC}$.
For that case, the minimum and maximum values among the $\sigma_\mathrm{DC}(\Theta_m)$ samples offer more information about the spread of the highly skewed DC conductivity distribution.

\begin{table}[]
    \centering
    \begin{tabular}{c|c|c|c|c}
         $q$ & $\overline{\sigma}_\mathrm{DC}$ & $\Delta \sigma_\mathrm{DC}$ & $\min_{\Theta_m}(\sigma_\mathrm{DC})$ & $\max_{\Theta_m}(\sigma_\mathrm{DC})$ \\\hline
         \SI{1.55}{\per\angstrom} & 0.66 & 0.13 & 0.39 & 1.2\\
         \SI{0.78}{\per\angstrom} & 230 & 1400 & 7.6 & 32000 \\
         both & 1.25 & 0.35 & 0.73 & 4.4
    \end{tabular}
    \caption{
    Statistical properties of DC conductivity distributions derived from the TDDFT-based collision frequency inferences shown in Fig.~\ref{fig:tddft-inference}.
    We report the mean (see Eq.~\eqref{eq:sigma_mean}), standard deviation (see Eq.~\eqref{eq:sigma_std}), minimum, and maximum DC conductivity in units of \SI{E+6}{\siemens\per\meter} corresponding to inferences using ELF data for $q=\SI{1.55}{\per\angstrom}$, $q=\SI{0.78}{\per\angstrom}$, and both wavenumbers simultaneously.
    }
    \label{tab:DCconductivity}
\end{table}

As expected given the disparate collision frequencies in Fig.~\ref{fig:tddft-inference}, the DC conductivity distributions corresponding to the different TDDFT-based inferences only barely overlap among each other.
Except for the very poorly constrained $q=\SI{0.78}{\per\angstrom}$ case, the distribution means $\overline{\sigma}_\mathrm{DC}$ do fall within the roughly 0.3\,--\,6 $\times$ \SI{E+6}{\siemens\per\meter} range of previously reported values for warm dense aluminum \cite{milchberg, sperling2015free, stanek_review_2024}.
Only the two-angle inference corresponding to Fig.~\ref{fig:tddft-inference}c produces DC conductivities near DFT-MD predictions of 1.3\,--\,2.6 $\times$ \SI{E+6}{\siemens\per\meter} \cite{stanek_review_2024}.

Overall, these findings demonstrate the difficulty of accurately constraining DC conductivity through an electron-ion collision frequency inferred from scattering data.
Not only do inferred values vary widely depending on the wavenumber or scattering angle used, but also a large range of values can maintain consistency with the original scattering spectrum.
Using data from multiple scattering angles may help mitigate but does not eliminate these challenges.

\section{Conclusions}
\label{sec:conclusions}
We have developed a method to infer dynamic collision frequencies and their uncertainties from electron loss functions, which are closely related to dynamic structure factors and observable XRTS spectra. Using a flexible parameterized form for the dynamic collision frequency, we apply Bayesian statistics and Monte Carlo methods to assess the range of collision frequencies that can reproduce ELFs generated by first-principles TDDFT. By exploring absolute and relative residuals and restricting the frequency range over which the objective function is evaluated, we find that the constraints on inferred collision frequencies are relatively local. Thus, the inferred values of the collision rate at frequencies near the plasmon peak will be much more reliable than values far from the peak. To some extent, this locality can be mitigated by simultaneously fitting different angles or wavenumbers that sample different frequency ranges. However, determining zero-frequency/DC conductivities for warm dense matter may require more focused methods of calculation \cite{andrade_negative_2018,melton_transport_2024,lin_electrical_2009} or measurement \cite{ofori-okai_dc_2024,chen_ultrafast_2021}.

\begin{acknowledgments}
We are grateful to Siegfried Glenzer, Patrick Knapp, William Lewis, and Michael MacDonald for helpful discussions and to Heath Hanshaw for pre-publication review.
All authors were partially supported by Sandia National Laboratories' Laboratory Directed Research and Development (LDRD) Project No.\ 233196.
This work was performed, in part, at the Center for Integrated Nanotechnologies, an Office of Science User Facility operated for the U.S. Department of Energy (DOE) Office of Science.
This article has been co-authored by employees of National Technology \& Engineering Solutions of Sandia, LLC under Contract No. DE-NA0003525 with the U.S. Department of Energy (DOE). The authors own all right, title and interest in and to the article and are solely responsible for its contents. The United States Government retains and the publisher, by accepting the article for publication, acknowledges that the United States Government retains a non-exclusive, paid-up, irrevocable, world-wide license to publish or reproduce the published form of this article or allow others to do so, for United States Government purposes. The DOE will provide public access to these results of federally sponsored research in accordance with the DOE Public Access Plan \url{https://www.energy.gov/downloads/doe-public-access-plan}.
\end{acknowledgments}

\appendix

\section{Born collision frequency sum rule}
\label{app:collision-sum-rule}

Here we derive the sum rule that constrains the width of the Born-like portion of our electron-ion collision frequency model, Eq.~\eqref{eq:collision-born}.
The general form of the Born collision frequency is
\begin{equation}
\nu_B(\omega) = \frac{-i}{6 \pi Z} \int_0^\infty dq \,q^6 S_\mathrm{i}(q) \frac{d\sigma}{d\Omega} \frac{\epsilon(q, \omega) - \epsilon(q, 0)}{\omega},
\label{eq:born}
\end{equation}
where $Z$ is the average ionization, $S_\mathrm{i}(q)$ is the ion structure factor, $d\sigma/d\Omega$ is the collision cross section, and $\epsilon(q, \omega)$ is the RPA dielectric function\cite{Thiele2008,Reinholz2000}.
For the simplest Born form of the collision frequency, we set \mbox{$S_\mathrm{i}(q) = 1$} and we use the Born cross section 
\begin{equation}
\frac{d\sigma}{d\Omega} = \frac{1}{4\pi^2}\left( \frac{4 \pi Z}{q^2}\right)^2.
\label{eq:born-crosssection}
\end{equation}
Under these simplifications, Eq.~\eqref{eq:born} becomes
\begin{equation}
\nu_B(\omega) = -\frac{2Z i}{3 \pi} \int_0^\infty dq \, q^2 \, \frac{\epsilon(q, \omega) - \epsilon(q, 0)}{\omega}.
\end{equation}
Since the static dielectric function $\epsilon(q, 0)$ is purely real \cite{ashcroft_solid_1976,mahan2013many}, the real part of the collision frequency is
\begin{equation}
\mathrm{Re}[\nu_B(\omega)] = \frac{2Z}{3 \pi} \int_0^\infty dq \, q^2 \frac{1}{\omega} \epsilon_2(q, \omega),
\label{eq:born-real}
\end{equation}
where $\epsilon_2(q, \omega)$ is the imaginary part of the dielectric function.

We are interested in a sum rule for the Born-like collision frequency, so we now integrate Eq.~\eqref{eq:born-real}:
\begin{align}
\int_0^\infty & d\omega \,\mathrm{Re}[\nu_B(\omega)] \nonumber\\
&= \frac{2Z}{3 \pi} \int_0^\infty dq \, q^2 \int_0^\infty d\omega \, \frac{1}{\omega} \epsilon_2(q, \omega).
\label{eq:sumrule_og}
\end{align}
The $\omega$ integral on the right side simplifies through the Kramers-Kronig relation for the dielectric function \cite{mahan2013many}:
\begin{equation}
\int_0^\infty d\omega \, \frac{1}{\omega} \epsilon_2(q, \omega) = \frac{\pi}{2}(\epsilon_1(q,0) - 1),
\label{eq:kk}
\end{equation}
where $\epsilon_1(q,0)$ is the real part of the dielectric function at $\omega = 0$.
Using the expression in Eq.~(15) of Ref.~\onlinecite{johnson2012thomson}, $\epsilon_1(q,0)$ can be expressed as
\begin{equation}
\epsilon_1(q,0) = 1 + \frac{4}{q^3 \pi} \int_0^\infty dp \, p \, \mathcal{F}(p) \ln\left|\frac{q^2 + 2pq}{q^2 - 2pq}\right| ,
\label{eq:rpa}
\end{equation}
where $\mathcal{F}(p)$ is the Fermi distribution as a function of electron momentum $p$.
Inserting Eqs.~\eqref{eq:kk} and \eqref{eq:rpa} into Eq.~\eqref{eq:sumrule_og}, we obtain
\begin{align}
& \int_0^\infty d \omega \, \mathrm{Re}[\nu_B(\omega)] \nonumber\\
&= \frac{4Z}{3 \pi} \int_0^\infty dp \, p \,\mathcal{F}(p) \int_0^\infty dq \, \frac{1}{q} \ln\left|\frac{q^2 + 2pq}{q^2 - 2pq}\right|.
\label{eq:sumrule_2}
\end{align}
Astonishingly, The $q$ integral is constant regardless of the value of $p$:
\begin{equation}
\int_0^\infty dq \, \frac{1}{q} \ln\left|\frac{q^2 + 2pq}{q^2 - 2pq}\right| = \frac{\pi^2}{2}.
\label{eq:q-integral-simple}
\end{equation}
Meanwhile, the integral over the Fermi factor is $k_BT \ln(1 + \exp(\mu / k_BT))$.
Thus, the sum rule for the real part of the simplest Born form of the collision frequency is
\begin{equation}
\int_0^\infty d \omega \, \mathrm{Re}[\nu_B(\omega)] = \frac{2\pi Z}{3} k_BT \ln\left(1 + e^{\mu / k_BT}\right).
\label{eq:sumrule-simple}
\end{equation}

If we replace the simple Born cross section of Eq.~\eqref{eq:born-crosssection} with a Born-Yukawa cross section of the form
\begin{equation}
\frac{d\sigma}{d\Omega} = \frac{1}{4\pi^2}\left( \frac{4 \pi Z}{q^2 + q_s^2}\right)^2,
\end{equation}
where $q_s$ is the inverse screening length, 
then the only change to Eq.~\eqref{eq:sumrule_2} is that the value of the $q$-integral becomes dependent on $p$:
\begin{align}
\int_0^\infty & dq \, \frac{q^3}{(q^2 + q_s^2)^2} \ln\left|\frac{q^2 + 2pq}{q^2 - 2pq}\right| \nonumber\\
& = \pi \tan^{-1}\left(\frac{2p}{q_s}\right) - \frac{\pi p q_s}{4p^2 + q_s^2}.
\label{eq:qint_yukawa}
\end{align}
As $q_s \rightarrow 0$, the right side of Eq.~\eqref{eq:qint_yukawa} becomes $\pi^2 / 2$ in agreement with Eq.~\eqref{eq:q-integral-simple}. 
For this cross section, the sum rule reads
\begin{align}
    \int_0^\infty & d \omega \, \mathrm{Re}[\nu_B(\omega)] \nonumber \\
     &= I(T, Z, \mu, q_s) \nonumber \\
    &= \frac{4Z}{3 \pi} \int_0^\infty dp \, p \, \mathcal{F}(p) \left(\pi \tan^{-1}\left(\frac{2 p}{q_s}\right) - \frac{\pi p q_s}{4 p^2 + q_s^2}\right) .
    \label{eq:sumrule-yukawa}
\end{align}
This integral is not analytically solvable except at $T = 0$, where the Fermi distribution becomes a step function. 
So, we numerically evaluate Eq.~\eqref{eq:sumrule-yukawa} for the system considered in the main text, solid-density aluminum at $k_BT=\SI{1}{\electronvolt}$.
Setting $Z=3$ and approximating $q_s= \sqrt{4 \pi n_e / T_\mathrm{eff}} = \SI{0.89}{\au}$ (where $n_e = \SI{0.027}{\au}$ is the electron density, $T_\mathrm{eff} = \max(k_BT, E_F)$ is the effective temperature, and $E_F$ is the Fermi energy), we obtain for Eq.~\eqref{eq:sumrule-yukawa} a value of $\SI{0.91}{\au}$ for $\mu = \SI{0.425}{\au}$ (pertaining to the ideal DOS calculations throughout most of the main text) and $\SI{0.62}{\au}$ for $\mu = \SI{0.321}{\au}$ (pertaining to the nonideal DOS calculations shown in Fig.~\ref{fig:tddft-inference-al-nonideal}).

Finally, we apply the Born-Yukawa version of the sum rule, Eq.~\eqref{eq:sumrule-yukawa}, to constrain the Born-like portion of our collision frequency model, Eq.~\eqref{eq:collision-born}.
Integrating Eq.~\eqref{eq:collision-born} gives
\begin{align}
    \int_0^\infty d \omega \, \nu_b(\omega)
    =& \int_0^\infty d \omega \, \frac{\nu_0}{1+(\omega/b_0)^{3/2}} \nonumber\\
    =& \frac{4\pi\nu_0b_0}{3\sqrt{3}}.
    \label{eq:sumrule-model}
\end{align}
To match the right sides of Eqs.~\eqref{eq:sumrule-yukawa} and \eqref{eq:sumrule-model}, we set the width $b_0$ to
\begin{equation}
    b_0 = \frac{3 \sqrt{3}}{4 \pi \nu_0} I(T, Z, \mu, q_s).
\end{equation}

\section{TDDFT simulations}
\label{app:tddft}

The real-time TDDFT calculations used to predict ELFs from first principles followed the same methodology described in earlier work \cite{baczewski_x-ray_2016,baczewski_predictions_2021,hentschel:2023}.
Briefly, these simulations included 32 aluminum atoms in a thermalized configuration obtained from an \emph{ab initio} molecular dynamics (MD) simulation.
The projector-augmented wave (PAW) method \cite{blochl_projector_1994} was used to explicitly model 3 valence electrons per ion, and exchange and correlation were treated with the adiabatic local density approximation (LDA) \cite{zangwill1980resonant,zangwill1981resonant}.
A plane-wave cutoff energy of \SI{500}{\electronvolt}, a time step of \SI{1}{\atto\second}, 5 electronic bands per aluminum ion, and Brillouin zone sampling with a $4\times 4 \times 4$ $\Gamma$-centered grid sufficed to converge dynamic response functions.
All first-principles calculations used a custom extension \cite{baczewski2014numerical,magyar2016stopping} of the Vienna \emph{Ab initio} Simulation Package (\textsc{VASP}) \cite{kresse1996efficient,kresse1996efficiency,kresse1999from}.

Dynamic response properties were obtained from the real-time electron density response $\delta n(r,t) = n(r,t) - n(r,0)$ to a sinusoidal probe potential with a Gaussian temporal envelope $V_\mathrm{probe}(r,t)$ \cite{baczewski_x-ray_2016}.
The probe's magnitude and duration were sufficiently small to remain in the linear-response regime, while the probe's wavevector was compatible with the supercell's periodicity and determined the accessed momentum transfer $q$.
The density-density response function $\chi(q,-q,\omega)$ relates the Fourier transforms of the density response $\delta n(q,\omega)$ and $V_\mathrm{probe}(q,\omega)$ so that the ELF is given by
\begin{align}
    \mathrm{Im}\left[-\frac{1}{\epsilon(q,\omega)}\right] =& -\frac{4\pi}{q^2} \mathrm{Im}[\chi(q,-q,\omega)] \nonumber\\
    =& -\frac{4\pi}{q^2}\mathrm{Im}\left[ \frac{\delta n(q,\omega)}{V_\mathrm{probe}(q,\omega)} \right].
    \label{eq:tddft-elf}
\end{align}

\begin{figure}
    \centering
    \includegraphics{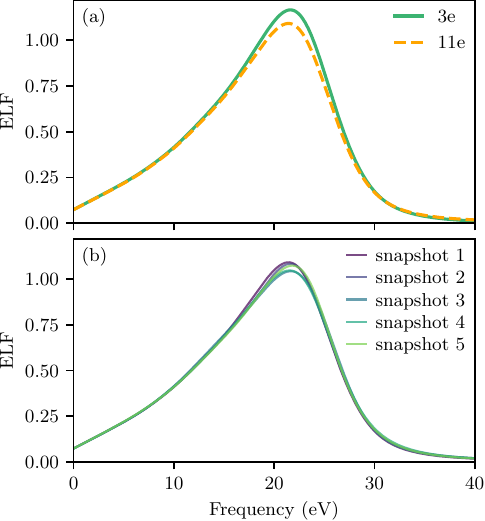}
    \caption{
    Sensitivities in the TDDFT ELF prediction for $q=\SI{1.55}{\per\angstrom}$.
    Panel (a) compares results computed with a 3-electron (3e) and an 11-electron (11e) PAW potential.
    Panel (b) compares results computed using five different atomic configurations sampled from an MD simulation.
    }
    \label{fig:tddft-uncertainties}
\end{figure}

Detailed interpretation of differences between TDDFT and Mermin-based predictions of the ELF relies on an understanding of the uncertainties in the TDDFT data.
Fig.~\ref{fig:tddft-uncertainties} highlights the dominant sensitivities relevant to the present work.
While the TDDFT data used to infer collision frequencies in the main text used a 3-electron PAW potential that pseudizes core and semi-core orbitals to isolate the free-electron response, explicitly modeling the $2s$ and $2p$ electrons through an 11-electron PAW potential results in a somewhat dampened plasmon peak (see Fig.~\ref{fig:tddft-uncertainties}a).
Additionally, we find minor variations in the height of the plasmon peak for different atomic configurations (see Fig.~\ref{fig:tddft-uncertainties}b) that are comparable in magnitude to the spread in the Mermin ELFs shown in Figs.~\ref{fig:tddft-inference}a and \ref{fig:tddft-inference-al-nonideal} of the main text.
However, the width of the plasmon peak along with the shape of its onset and decay are not sensitive to the details of the TDDFT calculation within the frequency range considered in the main text.

\bibliography{main.bib}

\end{document}